\documentclass[12pt,preprint]{aastex}

\shorttitle{Statistical Modeling of Solar Flare Activity} \shortauthors{Stanislavsky et al.}

\begin{document}

\title{Statistical Modeling of Solar Flare Activity from \\
Empirical Time Series of Soft X-ray Solar Emission}

\author{A. A. Stanislavsky,}
\affil{Institute of Radio Astronomy, National Academy of Sciences
of Ukraine,\\ 4 Chervonopraporna St., Kharkov 61002, Ukraine}
\email{alexstan@ri.kharkov.ua}

\author{K. Burnecki, M. Magdziarz, A. Weron}
\affil{Hugo Steinhaus Center, Institute of Mathematics and\\
Computer Science, Wroc{\l}aw University of Technology,\\ Wyb.
Wyspia\'{n}skiego 27, 50-370 Wroc{\l}aw, Poland}
\and

\author{K. Weron}
\affil{Institute of Physics, Wroc{\l}aw University of Technology,\\
Wyb. Wyspia\'{n}skiego 27, 50-370 Wroc{\l}aw, Poland}

\begin{abstract}

A time series of soft X-ray emission observed on 1974-2007 years (GOES) is analyzed. We
show that in the periods of high solar activity 1977-1981, 1988-1992, 1999-2003 the
energy statistics of soft X-ray solar flares for class M and C is well described by a
FARIMA time series with Pareto innovations. The model is characterized by two effects. One of them is a long-range
dependence (long-term memory), and another corresponds to heavy-tailed distributions. Their parameters are
statistically stable enough during the periods. However, when the solar activity tends to
minimum, they change essentially. We discuss possible causes of this evolution and
suggest a statistical model for predicting the flare energy statistics.

\end{abstract}

\section{Introduction}
\label{Introduction}

Individual solar cycles are different in form, amplitude and
length. At present, the accurate solar data is only available
for the most recent three cycles. Understanding the
long-term solar variability and predicting the solar activity is
an actual problem for solar physics. It is very important to
predict the time and strength of such events because these
disturbances can pose serious threats to man-made spacecrafts, can
disrupt electronic communication channels and can even set up huge
electrical currents in power grids (\citeauthor{Clark06}
\citeyear{Clark06}). It is enough to remind about serious problems
with GOES, Deuthsche Telecom, Telstars 401, etc. Satellite
operators would be glad to escape the unhappy surprise, and
mission planners are compelled to take into account the future
space weather forecast. Not only NASA satellites malfunctioned
because of the disturbances, but the global positioning system was
impaired. The cost to the airline industry arose as planes were
re-routed to lower altitudes, burning more fuel in force of
atmospheric drag.

As the geological records show, the Earth's climate has always
been changing. The reasons for such changes, however, have always
been subject to continuous discussions and are still not well
understood. In addition to natural climate changes the risk of
human influence on climate is seriously considered too. Any factor
that alters the radiation received from the Sun or lost to Space
will affect climate. So, \citeauthor{Mann98}(\citeyear{Mann98})
have clearly detected  a significant correlation between solar
irradiance and reconstructed Northern Hemisphere temperature. The
statistics indicates that during ``Maunder Minimum'' of solar
activity the climate was especially cold, but when the intensity
of solar radiance again increased from early nineteenth century
through to the mid-twentieth century, the period coincides with
the general warming. This, however, would either imply
unrealistically large variations in total solar irradiance or a
higher climate sensitivity to radiative forcing than normally
accepted. Therefore, other mechanisms have to be invoked. The most
promising candidate is a change in cloud formation because clouds
have a very strong impact on the radiation balance and because
only little energy is needed to change the cloud formation
process. According to satellite records taken from 1979 to 1992, 
Earth was 3\% cloudier during solar minima than at solar maxima
(\citeauthor{Svensmark97}\citeyear{Svensmark97}). One of the ways
to influence cloud formation might be through the cosmic ray flux
that is strongly modulated by the varying solar activity.
\citeauthor{Scafetta03}(\citeyear{Scafetta03}) argue that Earth's
short-term temperature anomalies inherit a L\'evy-walk memory
component from the intermittence of solar flares.

The aim of this paper is to present a statistical model for
predicting soft X-ray solar burst activity in the period of solar
cycle maxima.The paper is organized as follows. The random
features of solar activity is outlined in Section \ref{activity}.
The data set is described in Section \ref{flares}. In Section
\ref{predict} we present the essence  of our statistical
investigation of SXR flares. Interrelations
among statistical flare parameters, such as long-range dependence
index, Hurst exponent for X-ray flux and their evolution during
solar cycles, are analyzed. In the period of strong solar activity
the index of self-similarity is nearly constant (Section
\ref{results}). This feature can be used for predicting the power
of soft X-ray emission for the 24-th solar cycle near its solar activity maximum 
(2010-2014). The corresponding model is constructed in Section 
\ref{sectionempev}. Next, we give a summary and discussion of the 
main results. Finally, the conclusions are drawn in Section~\ref{conclusions}.

\section{Randomness in Solar Activity}\label{activity}

The solar 11-year cycle is driven by Sun's magnetic field. The
Sun's magnetic field is produced by a hydromagnetic dynamo process
underneath the solar surface and is cyclic in nature. This
fundamental theoretical idea was established by
\citeauthor{Parker55}(\citeyear{Parker55}). However, only within
the last few years, theoretical models of the solar dynamo have
become sophisticated enough to explain various aspects of the
solar activity. So, recently
\citeauthor{Gilman06}(\citeyear{Gilman06}) have made the first
attempt of using a theoretical dynamo model to predict the
strength of the upcoming cycle 24. They have shown that this cycle
will be the strongest in 50 years. But later
\citeauthor{Choudhuri07}(\citeyear{Choudhuri07}) pointed out that
some assumptions in the Dikpati-Gilman model are
unjustified. On the contrary, their model, based on the earlier work of
\citeauthor{Nandy02}(\citeyear{Nandy02}), predicts that the cycle 24 will be
weaker than the 23-rd. The key problem here is the following:
the dominant processes like the magnetic field
advection and toroidal field generation by differential rotation are fairly 
regular during the rising phase of a cycle from a minimum to a
maximum, and hence a good knowledge of magnetic configurations during a
minimum would enable a good theoretical model to predict the next
maximum reliably. However, the dominant process in the declining
phase of a cycle contains the poloidal field generation by the
Babcock-Leighton mechanism which involves randomness (primary
cause of solar cycle fluctuations) and cannot be predicted in
advance by any deterministic model. That is why, although active
regions appear in a latitude belt at a certain phase of the solar
cycle, where exactly within this belt the active regions appear
seems random. Since the poloidal field generated from an active
region depends on the tilt, the scatter in the tilts introduces
randomness in the poloidal field generation process.

The other feature of solar activity is that there is a ``magnetic
persistence'' between the surface polar fields and spot-producing
toroidal fields, generated by differential rotation shearing
(\citeauthor{Dikpati06}\citeyear{Dikpati06}). This means that the
Sun retains a memory of its magnetic field for a long time (about
20 years or so). The solar cycle prediction is similar to that
employed in global atmospheric dynamics over the last ten years.
Such models predict changes in certain global characteristics of a
cycle, without attempting to reproduce details that occur on
smaller spatial scales and shorter time scales. The interrelation
between global characteristics and small scale processes is an
open problem, and meanwhile some effects of smaller scales are
included in parametric form.

\section{X-ray Flare Observations}\label{flares}

Solar activity is a many-sided phenomenon. It includes flares, prominence eruptions,
coronal mass ejections, solar energetic particles, various radio bursts, high-speed solar
wind streaming from coronal holes, etc. Solar flares are the most energetic and violent
events occurring in the solar atmosphere. The energy release in a flare ranges from
10$^{26}$ to 3$\times$10$^{32}$ ergs. Magnetic reconnection is considered to play a
central role in any flare energy release.

Observations of solar flare phenomena in X-rays became possible in
the 1960s with the availability of space-borne instrumentation.
Since 1974 broad-band soft X-ray emission of the Sun has been
measured almost continuously by the meteorology satellites
operated by NOAA so as the Synchronous Meteorological Satellite
(SMS) and the Geostationary Operational Environment Satellite
(GOES). The first GOES was launched by NASA in 1975, and the GOES
series extends to the currently operational GOES 11 and GOES 12.
From 1974 to 1986 the soft X-ray records are obtained by at least
one GOES-type satellite; starting with 1983, data from two and
even three co-operating GOES are generally available. The X-ray
sensor, part of the space environment monitor system aboard GOES,
consists of two ion chamber detectors, which provide whole-sun
X-ray fluxes in the 0.05-0.3 and 0.1-0.8 nm wavelength bands.
Solar soft X-ray flares are classified according to their peak
burst intensity measured in the 0.1-0.8 nm wavelength band by
GOES. The letters (A, B, C, M, X) denote the order of magnitude of
the peak flux on a logarithmic scale, and the number following the
letter gives the multiplicative factor, i.e., A$n=n\times
10^{-8}$, B$n=n\times 10^{-7}$, C$n=n\times 10^{-6}$, M$n=n\times
10^{-5}$ and X$n=n\times 10^{-4}$ W/m$^2$. In general, $n$ is
given as a float number with one decimal (prior to 1980, $n$ is
listed as an integer). No background subtraction is applied to the
data. Now the data is widely available from the NOAA Space
Environment Center site
(http://www.ngdc.noaa.gov/stp/SOLAR/ftpsolarflares.html).

In the meantime, a wealth of data has been accumulated. It makes worthwhile
re-investigating the temporal and spatial features of soft X-ray (SXR) flares on an
extensive statistical basis. \citeauthor{Li98} (\citeyear{Li98}) studied the distribution
of the X-ray flares (M$\geq$1) from 1987 to 1992 with respect to helio longitude. They
have shown that the flares were not uniformly distributed in longitude. The temporal
analysis of X-flare statistics concerns basically the waiting-time distribution (see, for
example, \citeauthor{Boffetta99} \citeyear{Boffetta99}; \citeauthor{Moon01}
\citeyear{Moon01}; \citeauthor{Lepreti01}\citeyear{Lepreti01};
\citeauthor{Weatland02}\citeyear{Weatland02}; \citeauthor{Veronig02} \citeyear{Veronig02}
and so on). In the present analysis we make use of SXR flares observed by GOES during
1976-2006. Our consideration will be devoted only to the energy statistics of soft X-ray
solar flares in time.

\section{Predictive Tool Description}\label{predict}

Our analysis is based on the properties of fractional autoregressive integrated moving average (FARIMA) processes 
(\citeauthor{Beran94} \citeyear{Beran94}). They are widely used in modeling of 
various complex physical systems. The FARIMA($p$,$d$,$q$) process is defined
as the solution of the equation $\Phi(B)\Delta^dX(n)=\Theta(B)\epsilon_n$, $n\in{\bf Z}$, where
$B$ is the shift operator $BX(n)=X(n-1)$ and $\Delta$ is the difference operator, i.\ e. 
$\Delta X(n)= X(n)-X(n-1)$. Here $\Phi$ and $\Theta$ are the polynomials of degree $p$ and $q$
respectively, $d$ takes fractional values, either positive or negative, and ``innovations'' 
$\epsilon_j$ are independent and identically distributed (i.i.d.) 
random variables. The polynomials $\Phi$ and $\Theta$ correspond to autoregressive (AR) 
and moving average (MA) parts, respectively. The linear representation of FARIMA processes 
takes the form
\begin{equation}
X(n)=\sum^\infty_{j=0}c_{n-j}\epsilon_j\,,\label{eq1}
\end{equation}
for details see (\citeauthor{Beran94} \citeyear{Beran94}). The innovations may be
either Gaussian, non-Gaussian with finite variance or they may have infinite variance.
For infinite variance innovations $\epsilon$, one may consider, for example, symmetric
and skewed stable distributions, as well as Pareto distributions. Both are characterized
by the parameter $\alpha$ and their tails $P(\epsilon > x)$ satisfy
\begin{equation}
P(\epsilon>x)=1-F(x)\sim x^{-\alpha},\qquad {\rm as}\qquad x\to\infty\label{eq2},
\end{equation}
where $F(x)$ denotes the corresponding distribution function and $\sim$ denotes that 
the ratio of the left-hand side to the right-hand one tends to 1, as $x\to\infty$. 
It should be noted that the L\'evy-stable distributions have $0<\alpha<2$ whereas for the Pareto
distribution the parameter $\alpha$ is greater than zero. The resulting process $X(n)$ will be 
long-range dependent and L\'evy-stable if the innovations are L\'evy-stable, and asymptotically
will be in the domain of attraction of a L\'evy-stable distribution if the innovations are Pareto 
(see \citeauthor{Samorod94} \citeyear{Samorod94}). Moreover, such FARIMA processes are asymptotically 
self-similar with $d-1/\alpha$. 

The L\'evy-stable distribution, named after
the French mathematician Paul L\'evy who investigated the behavior of sums of independent random variables, 
is most conveniently described by its characteristic function $\phi(\theta)$ -- the inverse Fourier transform 
of the probability density function. The most popular form of the characteristic function of a 
L\'evy-stable random variable is given by the expression
\begin{equation}
\log\phi(\theta) = \cases{ 
  -\sigma^\alpha|\theta|^\alpha\{1-i\beta\,{\rm sign}(\theta)\tan(\pi\alpha/2)\}+i\mu \theta,\qquad \alpha\neq 1,\cr
  -\sigma\,|\theta|\,\{1+2i\beta\,{\rm sign}(\theta)\log|\theta|/\pi\}+i\mu \theta,\quad\qquad \alpha = 1,}
\end{equation}
where $0<\alpha\leq 2$, $-1\leq\beta\leq 1$, $\sigma>0$ and $\mu \in {\bf R}$ are parameters of this distribution (\citeauthor{Samorod94} \citeyear{Samorod94}). The Pareto distribution, 
introduced by the Italian economist Vilfredo Pareto, is a power law probability density that we represent 
in the form of $f(x)=\alpha\lambda^\alpha(\lambda+x)^{-\alpha-1}$, where $\lambda$ and $\alpha$ are positive constants
(\citeauthor{Burn05} \citeyear{Burn05}).

The power-law behavior of the tails implies that the variance is infinite if $\alpha<2$. The
tail index (exponent) $\alpha$ controls the rate of decay of the
tail of the distribution function $F$. Modeling with FARIMA time series with infinite 
variance allows to take into account heavy tails. Through a suitable choice of coefficients 
$c_{n-j}$ one can also add long-term memory effects. The FARIMA processes is an useful family 
of models because it offers a lot of flexibility in modeling long-range and short-range
dependence by choosing the memory parameter $d$ and appropriate autoregressive and moving
average coefficients in expression (\ref{eq1}).

The problem of estimating the exponent in heavy-tailed data has a
long history in statistics because of its practical importance.
The presence of heavy tails in data was firstly noted in the work
of \citeauthor{Zipf32}(\citeyear{Zipf32}) in his study of word
frequencies in languages. Next,
\citeauthor{Mandelbrot60}(\citeyear{Mandelbrot60}) noted their
presence in financial data. Since the early 1970s the heavy-tailed
behavior has been noted in many other scientific fields (see, for
example, reviews of \citeauthor{Adler98} \citeyear{Adler98};
\citeauthor{Park00} \citeyear{Park00}). However, the availability
of huge amount of various data poses a set of new challenges for
the problem of estimating the tail index. The point is that the
data can be contaminated by extraneous oscillations, different
noises with finite variance and so on. This makes the analysis of 
heavy-tailed data more complicated (\citeauthor{Janicki94}\citeyear{Janicki94}; 
\citeauthor{Lynch05} \citeyear{Lynch05}). The time series of soft 
X-ray solar emission relates to such problematic data
(\citeauthor{Baiesi06}\citeyear{Baiesi06}). Therefore, for
reliability we will estimate the tail index by different
statistical tests.
One of them is based on the asymptotic {\it max self-similarity}
properties of heavy-tailed maxima
(\citeauthor{Stoev06}\citeyear{Stoev06}). In this test the maximum
values of data are calculated over blocks of size $m$, scaled at
rate of $m^{1/\alpha}$. By examining a sequence of growing block
sizes $m=2^j$, $1\leq j\leq\log_2 N, j\in{\bf N}$, and
subsequently estimating the mean of logarithms of block-maxima one
obtains an estimation of the tail index $\alpha$. Another
estimator, that we use, under the assumption of the L\'evy stable law 
applies the \citeauthor{McCulloch86}(\citeyear{McCulloch86}) quantile fit.

\section{Cycling of Self-similarity}\label{results}

\begin{figure}
\begin{center}
\includegraphics[width=11.cm]{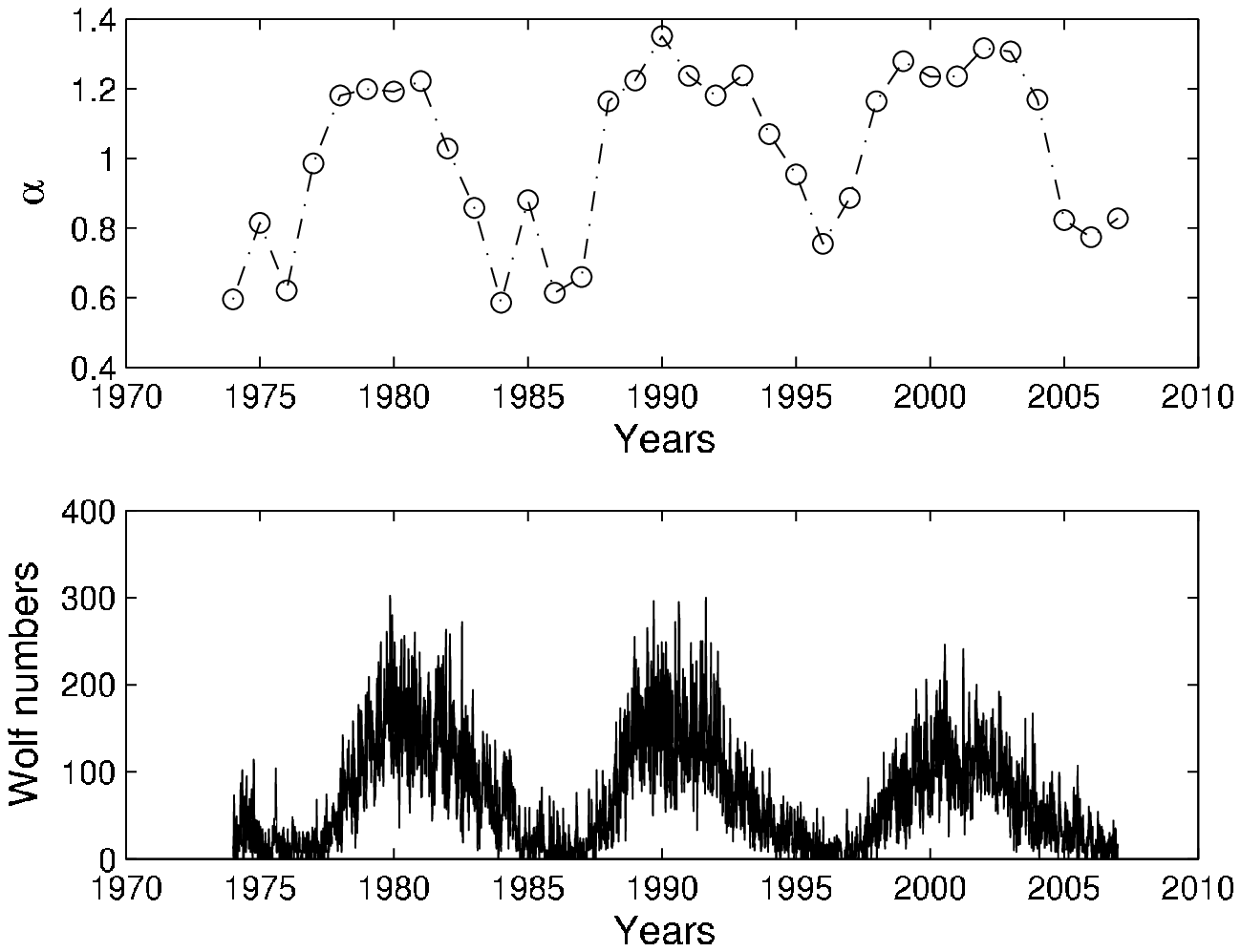}
\caption{The evolution of the tail index $\alpha$ during the last solar cycles 1974-2006
(top picture) obtained via the max self-similarity method, the bottom picture shows Wolf numbers (characterizing solar activity) in
this period.} \label{maxspec}
\end{center}
\end{figure}

Using the max self-similarity estimator and considering our
data in year intervals, we have analyzed how the solar cycling
influences on the tail index. Figure \ref{maxspec} shows a clear
correlation between solar activity and the index. When the solar
activity is around maxima, the tail index is larger than one,
whereas in minima it tends to fall down less than one. The index
value in the period of high solar activity almost coincides with
the result of \citeauthor{Weron05} (\citeyear{Weron05}). Although
their data insert only X-ray solar bursts of C and M type, the
value $\alpha$ was estimated to be $1.2674$. The present analysis
extends the tail index analysis on some cycles and speaks surely
that the index tendency observed earlier is kept at least
during the recent three solar cycles.

The McCulloch's testing (\citeauthor{McCulloch86}\citeyear{McCulloch86}) gives similar results.
This test shows that the tail parameter $\alpha$ 
depends on the solar activity value during the three solar cycles. Of course, the estimator has a 
particular character. It is convenient for the analysis of random variables with a stable 
distribution. Nevertheless, the distribution also has heavy tails. Our first aim is to find 
such distribution that will be best fitted to the experimental series.

\section{Solar flares}
\label{sectionempev}

\begin{figure}
\begin{center}
\includegraphics[width=15.cm]{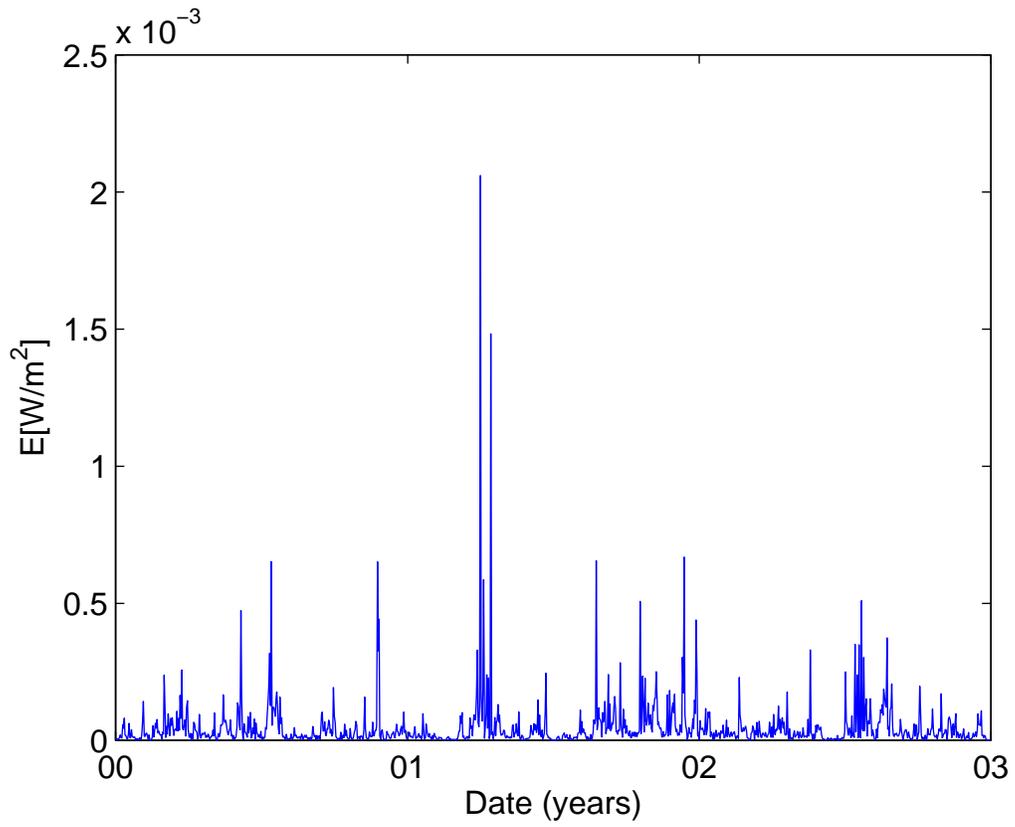}
\end{center}
\caption{Energy-time series of solar flares from 2000 January 01 to 2002 December
31.} \label{data}
\end{figure}

Now we restrict our attention to such time intervals, in which the solar activity is
strong (in particular, 1978-1981, 1988-1992 and 1999-2003). We use X-ray flare data from
GOES satellite, that contain information about time of appearance and energy of solar
flares (from http://www.ngdc.noaa.gov/stp/SOLAR/ftpsolar-flares.html). The captured
energy was transmitted by X-rays emitted during blasts on a solar surface from 2000
January 1 to 2002 December 31. We aggregated the energy values on a daily basis. The time
series is presented in Fig.~\ref{data}.

The first estimation procedure of the self-similarity parameter
$H$ (i.\ e. the Hurst exponent)is the so-called finite impulse response transformation
(FIRT).  The FIRT estimator involve an array of coefficients. The
array is made out of finite impulse response coefficients. The
estimator $H_{FIRT}$ is obtained by performing a log-linear
regression on the coefficients and measuring the slope
(\citeauthor{SP}\citeyear{SP}). It is important to note that the
estimator $H_{FIRT}$ is unbiased for all $\alpha$ falling in the range $(0,2)$.

An alternative method of testing scaling and correlation
properties of a time series is the variance of residuals
method (VR) (\citeauthor{peng} \citeyear{peng}). First, the series
is divided into blocks of size $m$. Then, within each block, the
partial sums of the series are calculated. A least­-squares line
is fitted to the partial sums within each block, and the sample
variance of the residuals is computed. The variance of residuals
is proportional to $m^{2H}$. This variance of residuals is
computed for each block, and the median (or average) is computed
over the blocks. A log­-log plot versus $m$ should follow a straight
line with a slope of $2H$.

The R/S method is one of the oldest and better known methods for estimation of the Hurst parameter (\citeauthor{Hurst51}\citeyear{Hurst51}). Hurst found that drought in the Nile Valley is not a random phenomenon, but rather that the region is inclined to become progressively more arid after a succession of long droughts. It is widely known that for Gaussian (i.e. finite variance) time series the method returns $H$ (\citeauthor{ManWal69}\citeyear{ManWal69}). The popularity of the method has been also a source of misunderstandings and errors. This is due to the fact that, in general, for power-law distributed time series (i.e. with infinite variance) the method yields $d+1/2$. In particular for Pareto or L\'evy stable distributions the output is $H-1/\alpha+1/2$  (\citeauthor{TaqTev96}\citeyear{TaqTev96}). The method is based on R/S (rescaled adjusted range) statistics. The series is divided into blocks. Then, within each block, the statistics is calculated. Finally, arithmetic means of the values of the statistics over the blocks are calculated and a least squares line is fitted to the mean for different lengths of the blocks. The slope should be equal to $d+1/2$.

For the finite variance cases, the interpretation of the FIRT and VR estimators is very
similar to the Hurst exponent: if only short-range correlations (or no correlations at
all) exist in the studied series, then $H_{FIRT}=H_{VR}=1/2$; if there is a correlation
then $H_{FIRT}=H_{VR}\neq 1/2$. Moreover, if the estimator $H_{FIRT}=H_{VR}$ is greater
than $1/2$, the time series is persistent and if $H_{FIRT}=H_{VR}<1/2$, then the time
series is not persistent.

Note that both estimators give an information on memory and not on distribution of the
process increments.

\begin{figure}
\begin{center}
\includegraphics[width=11.cm]{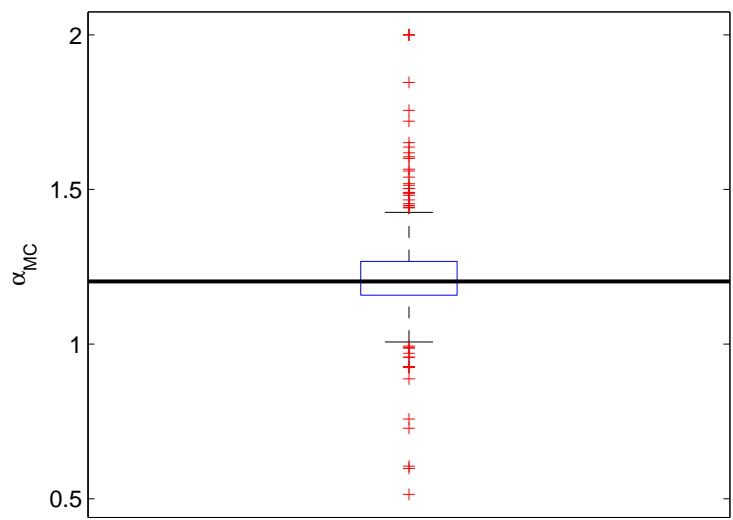}
\end{center}
\caption{Values of the calculated $\alpha$ estimators for the simulated FARIMA time series. Solid line represents the value of the estimator for the analyzed data.}\label{fig_alpha}
\end{figure}

The analysis of the data shows that the tails of the underlying
distribution conform to the power law. Hence, we model the data by a FARIMA process with Pareto innovations. As the power-law distributions belong to the domain of attraction of stable law (see, eg. \citeauthor{Janicki94}\citeyear{Janicki94}) the resulting distribution of the FARIMA process should be close to the stable one. We applied
the McCulloch quantile fit to obtain the parameters of the
distribution (\citeauthor{McCulloch86}\citeyear{McCulloch86}). The
value of $\alpha$ was estimated to be $1.213$, see Figure \ref{fig_alpha}. One may check that the estimated value of $\alpha$ of simulated FARIMA times series with Pareto innovations is usually underestimated, see e.~g. Figure \ref{fig_alpha}. Therefore, we assume that the innovations in our model follow the Pareto law with $\alpha = 1.25$.

According to \citeauthor{Weron05} (\citeyear{Weron05}), in order
to recover both the self-similarity exponent $H$ and the memory
parameter $d$ (hence, the distribution parameter $\alpha$) we can
use the following BMW$^2$ computer test. The surrogate data are
obtained here by a random shuffling of the original data
positions.

\begin{itemize}
\item If the process is FARIMA with Gaussian noise, then the values of the estimator should
change to $1/2$ for the surrogate data independently on the initial values.
\item If the process is FARIMA with $\alpha$-stable or Pareto noise for $\alpha<2$, then the values of the estimator should change to $1/\alpha$ for the surrogate data independently on the initial values.
\end{itemize}

The above formalism can be easily applied to determine basic
features of an empirical data series. Now we employ this to
study an empirical time series recorded from the
system describing the energy of solar flares (Fig. \ref{data}). The obtained values
of the parameters are listed in Table~\ref{tab1}. Therefore, from
the results for the surrogate data, the corresponding estimates
for the parameter $1/\alpha$ are:
$1/\alpha_{FIRT}=H_{FIRT}=0.8452$ and
$1/\alpha_{VR}=H_{VR}=0.7722$. We observe that the estimators are
close to the one assumed in our model: $1/\alpha_{MC}=0.8$. Moreover, we choose $d = 0.19$ as the highest admissible value of $d$, which is close to the one obtained via the R/S method for the original data, see Table \ref{tab1}.

\begin{figure}
\begin{center}
\includegraphics[width=11.cm]{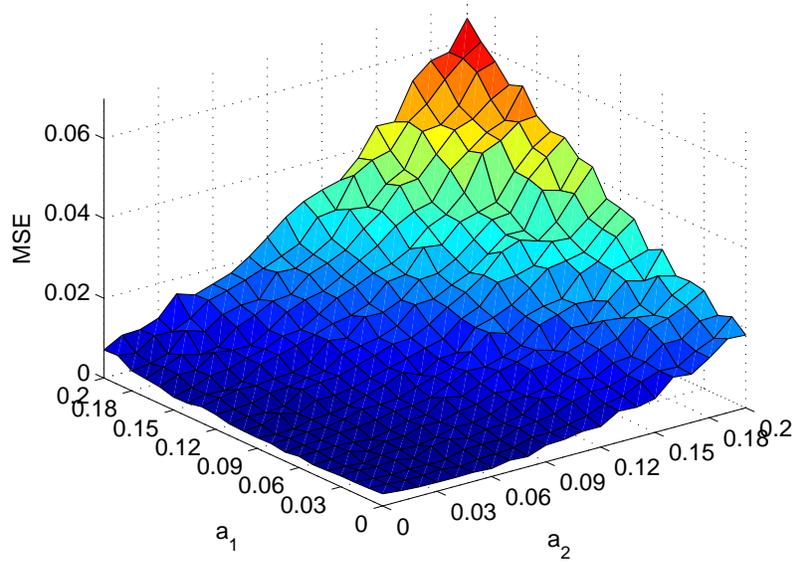}\\
\end{center}
\caption{Mean squared error on the basis of the calculated $H$ and $d$ estimators for the simulated FARIMA time series with respects to the estimators obtained for the analyzed data for different $a_1$ and $a_2$.}\label{fig_mse}
\end{figure}

One may notice that the estimators of $H$ obtained via FIRT and VR methods, see Table \ref{tab1}, are greater than theoretically admissible in the FARIMA model, i.\ e. they exceed one. As stated in \citeauthor{Burn08} (\citeyear{Burn08}), this can be justified by performing simulations of the FARIMA processes and estimating the parameter $H$ on the simulated time series via different methods. It appears that a reasonable percentage of the values of $H$ is higher than the data estimates, so one can not reject the hypothesis that the underlying model is the FARIMA(0,$d$,0) process. Nevertheless, we now look for an enhanced model which would describe better the behavior of different estimators obtained for the original and shuffled data, and would improve the fit in terms of the prediction error. Thus, we propose a slight generalization of the model incorporating the short-dependence component, namely FARIMA (2,$d$,0) model. We estimated the AR(2) coefficients: $a_1$ (linear term) and $a_2$ (quadratic term) via the mean-square error (MSE) minimalization scheme taking into account three statistics: FIRT, VR and RS, see Figure \ref{fig_mse}. The estimated values are: $a_1 = 0.02$ and $a_2 = 0.03$. The FARIMA processes were generated
according to the algorithm presented by \citeauthor{Stoev04}(\citeyear{Stoev04}).

\begin{figure}
\begin{center}
\includegraphics[width=8.cm]{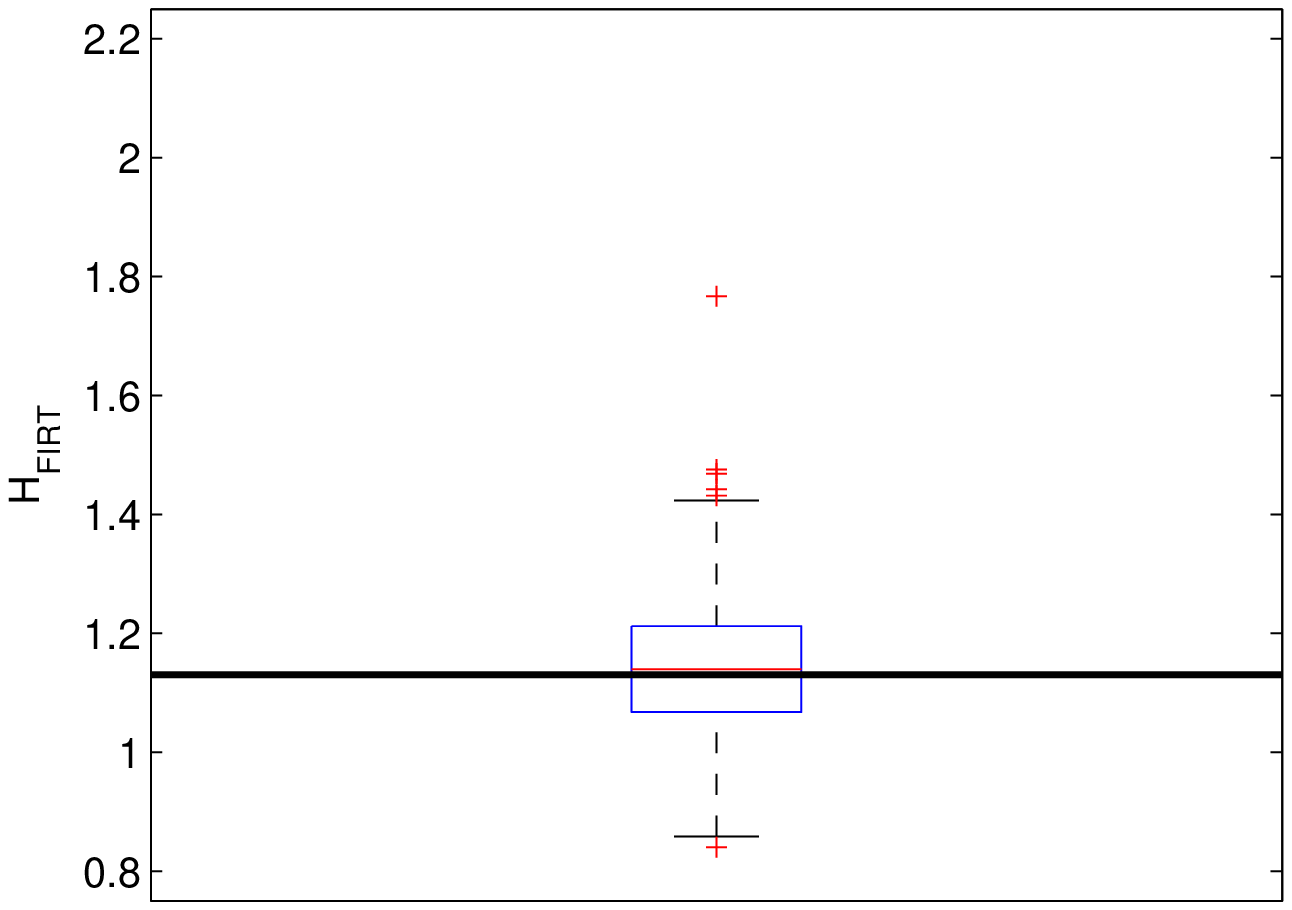}\includegraphics[width=8.cm]{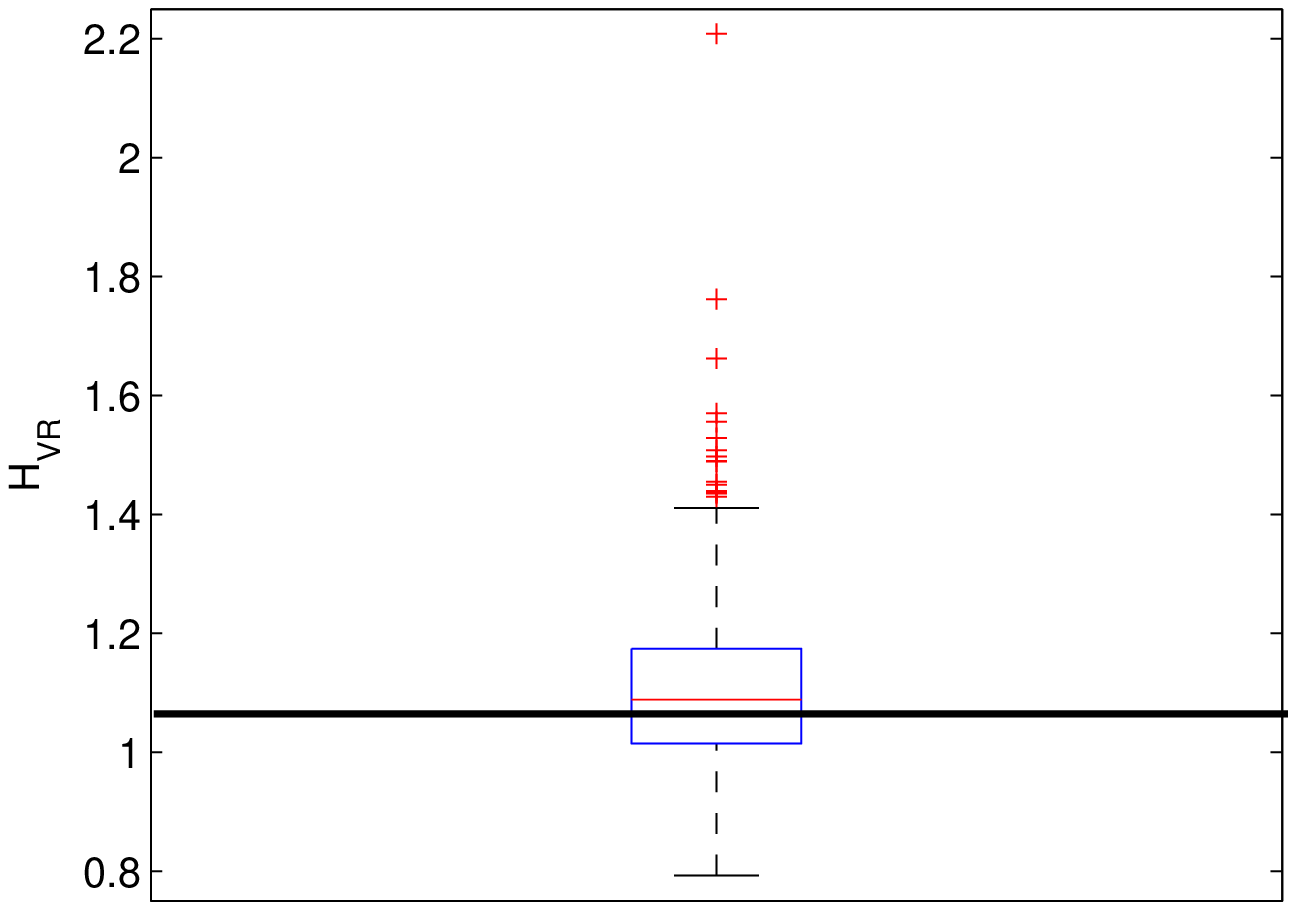}\\
\includegraphics[width=8.cm]{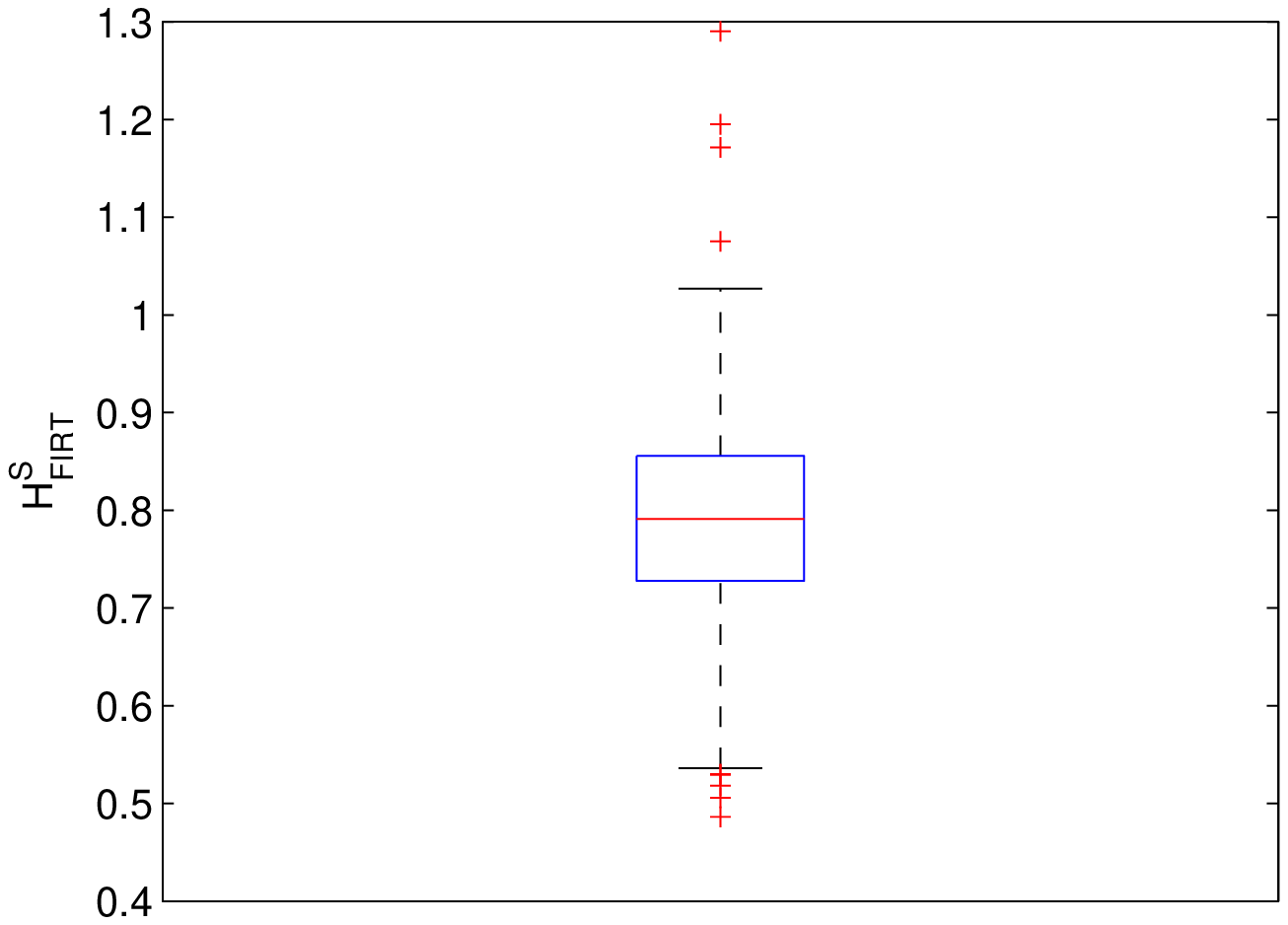}\includegraphics[width=8.cm]{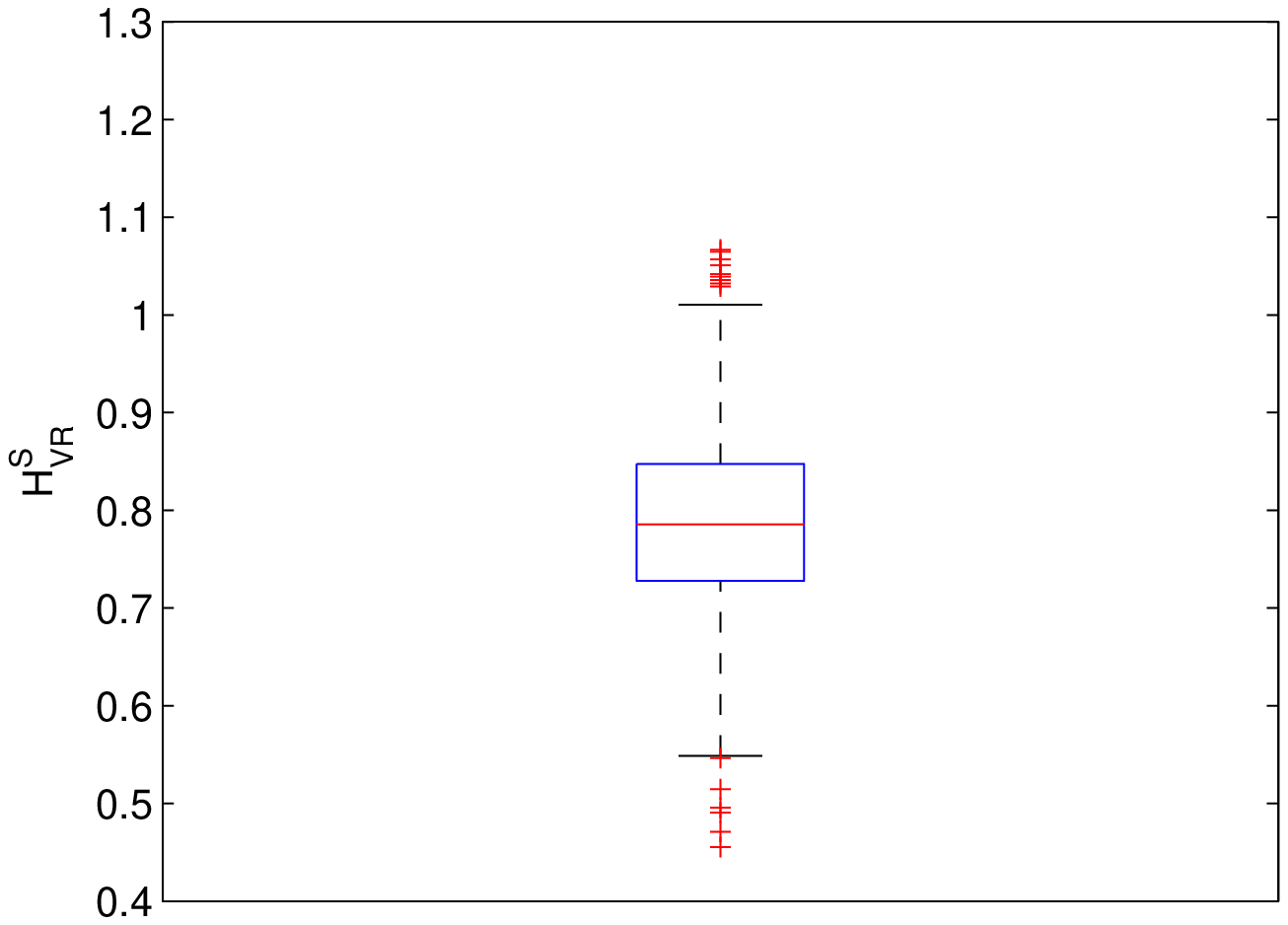}
\end{center}
\caption{Values of the FIRT and VR estimators for the simulated time series (top
panel) and the surrogate data (bottom panel) for the generated FARIMA processes. Solid line represents the value of the estimator for the analyzed data.}\label{fig_box_est}
\end{figure}

\begin{figure}
\begin{center}
\includegraphics[width=8.cm]{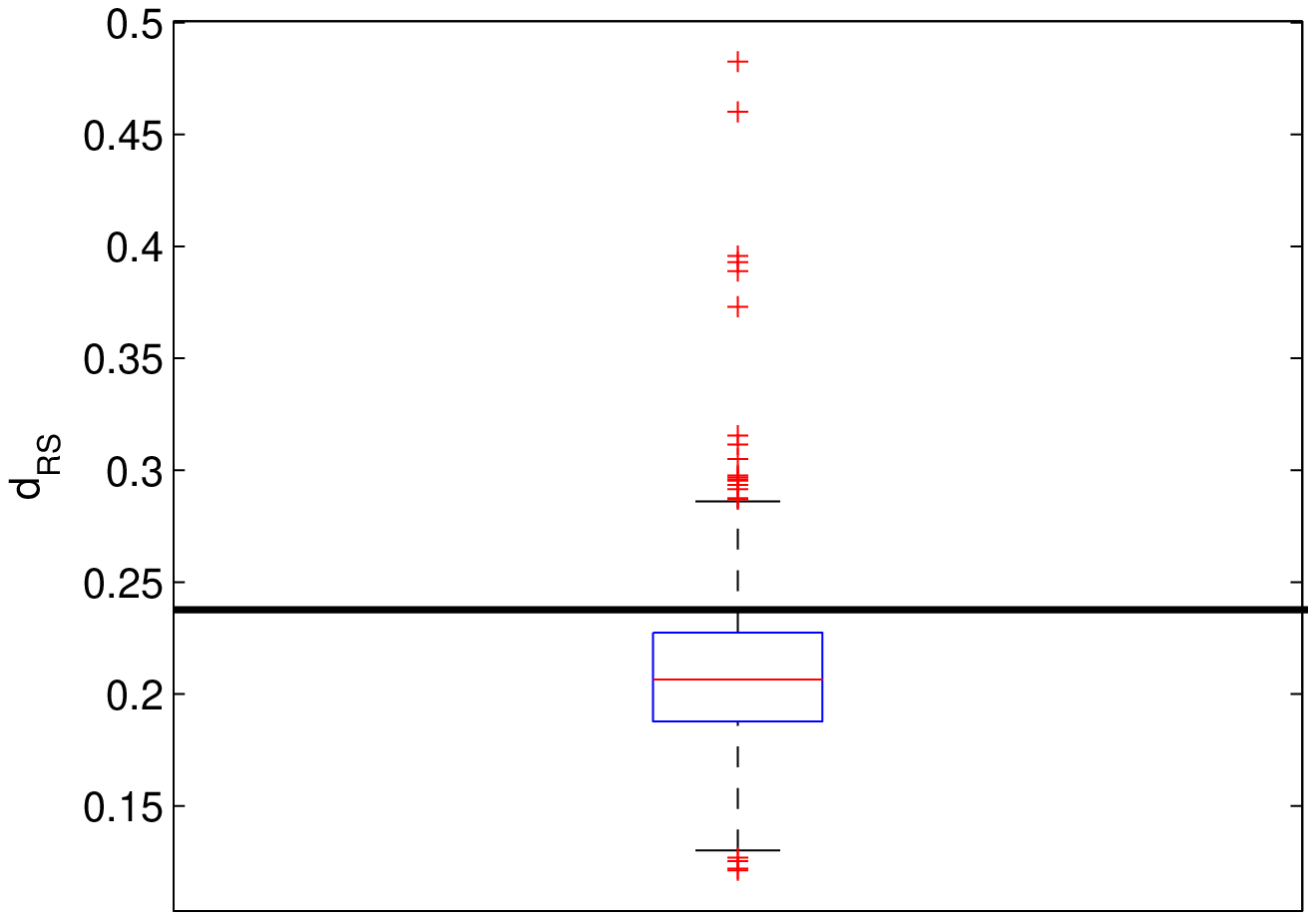}\\
\includegraphics[width=8.cm]{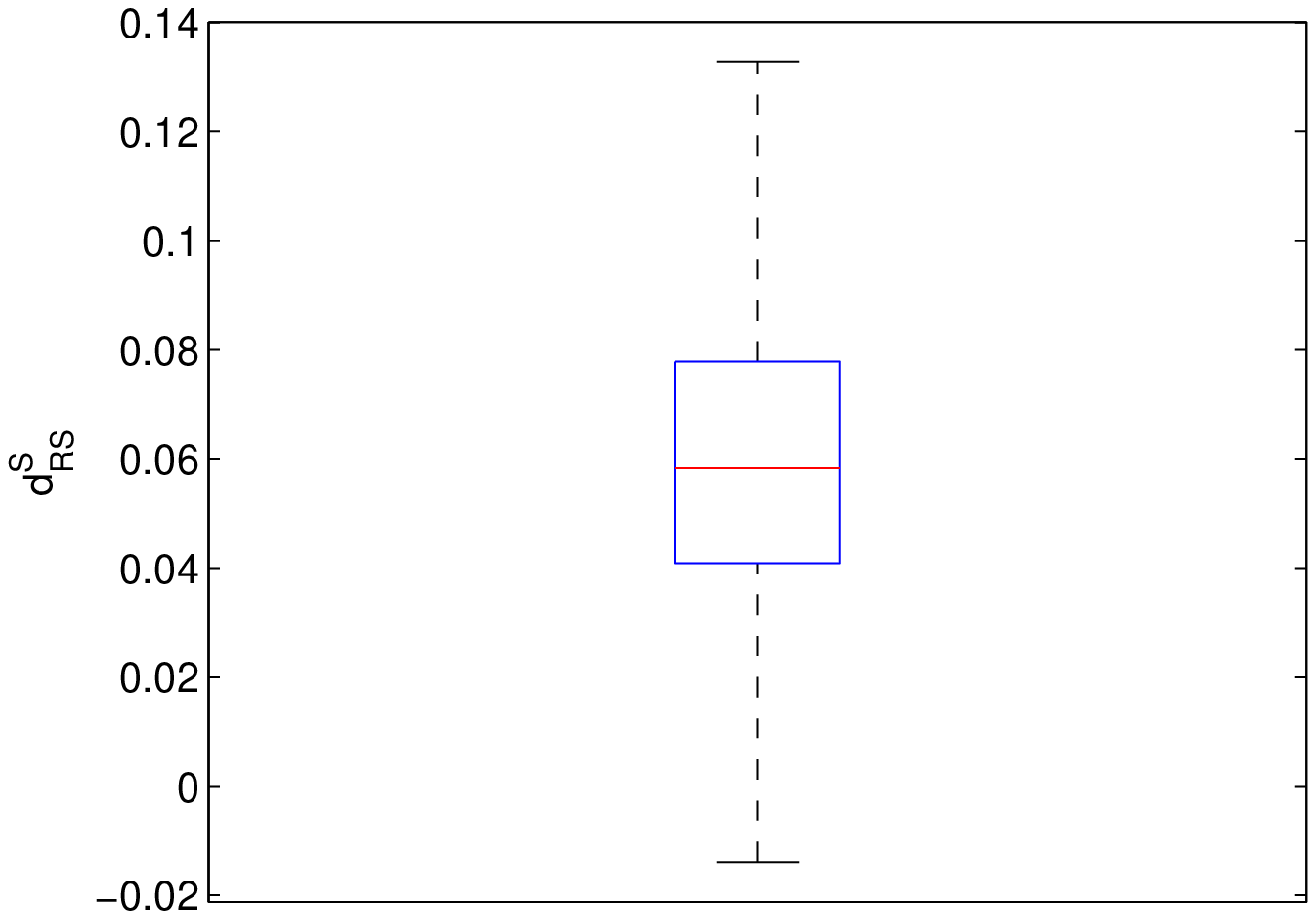}
\end{center}
\caption{Values of the RS estimator $H_{RS}$ for the simulated time series (top panel)
and the surrogate data (bottom panel) for the generated FARIMA processes. Solid line represents the value of the estimator for the analyzed data.}\label{fig_box_rs}
\end{figure}

We calculate $H_{FIRT}$, $H_VR$ and $d_{RS}$ estimators for the
simulated FARIMA (2,$d$,0) processes (top
panel in Figs. \ref{fig_box_est} and \ref{fig_box_rs}) and the corresponding shuffled data (bottom panel in Figs. \ref{fig_box_est} and \ref{fig_box_rs}). We generate $1000$ trajectories of
size $2^{10}$ which is close to the length of the original solar data, i.\ e. 1089 and
present the results in form of the so-called box plots.  The
box plot has lines at the lower quartile, median, and upper quartile
values. The whiskers are lines extending from each end of the box
to show the extent of the rest of the data. Outliers are data with
values beyond the ends of the whiskers. If there is no data
outside the whisker, a dot is placed at the bottom whisker. The results
correspond to the analysis of the original solar data included in
Table~\ref{tab1}. Thus, our FARIMA  simulations reconstruct 
well the structure of the original solar flares data. 

We conclude that the proper model could be based on the FARIMA(2,$d$,0) process with Pareto innovations with
the parameters $a_1 = 0.02$, $a_2 = 0.03$, $d=0.19$, and $\alpha = 1.25$ which has the
long-range dependence property since $d>1-2/\alpha$ (\citeauthor{Burn08} \citeyear{Burn08}). The Pareto distribution is also convenient for the present analysis
because it gives a description of positive random variables whereas the L\'evy-stable one for $1<\alpha<2$ is related to
both positive and negative random variables, but a time series of x-ray flare energy is quite positive by definition.

\begin{figure}
\begin{center}
\includegraphics[width=11.cm]{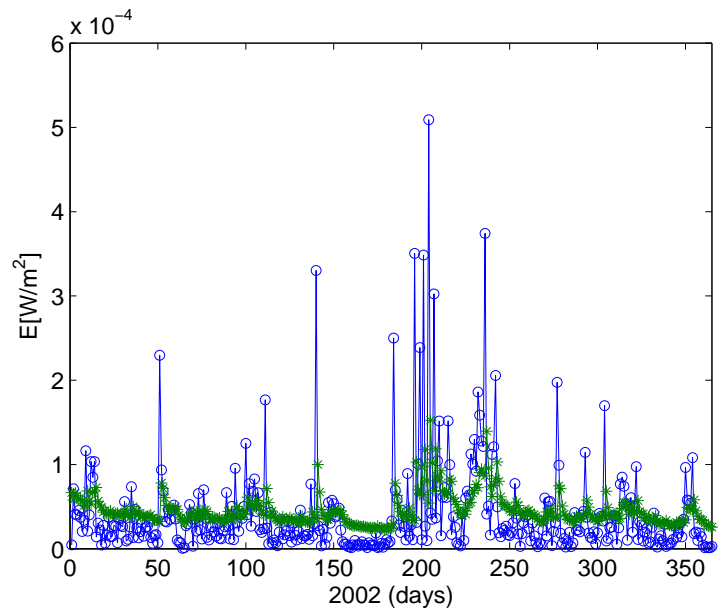}
\end{center}
\caption{Solar flare data and prediction in the FARIMA(2,d,0) model.}\label{fig_pred}
\end{figure}

We also calculated the 1-day-ahead prediction for the FARIMA(2,$d$,0) time series (for the prediction discussion in the infinite variance FARIMA case see \citeauthor{Kokoszka95}\citeyear{Kokoszka95}). The results are depicted in Figure \ref{fig_pred}.

\section{Concluding Remarks and Discussion}
\label{conclusions}

In this paper we demonstrate how self-similar models driven by
L\'evy stable noise can be useful for modeling X-ray solar data.
To be more precise we have suggested the FARIMA(2,$d$,0) model with
$\alpha$-stable noise for predicting solar flare appearance in the
period of a strong solar activity. 

The procedure is illustrated in Section \ref{sectionempev} for the
captured energy transmitted by X-rays emitted during blasts on a solar surface from 2000
January 1 to 2002 December 31. Comparing the values of the different estimators for the
original data series and for the surrogate data we estimated the components of the
self-similarity index corresponding to the memory of the time series ($d$) and to the
tail properties of the time series values distribution ($\alpha$), see Table \ref{tab1}.
Thus, this allows in principle to build a proper physical model for analyzing the
solar activity.

The analysis of soft X-ray emission observations shows that this
series is enough complicated in nature. We have seen a strong
dependence of statistics on solar cycling in the data. However, if
one takes year intervals, then this cycle influence becomes not so
strong. This allows one to reconstruct a statistical model for
predicting the soft X-ray solar activity on the nearest solar
cycle. The soft X-ray solar time series contains both long-range
dependence and heavy-tailed effects. The first creates a
random number of strong bursts on a background, and the second
forms their persistence between each other. The most convenient
model for their joint description is a FARIMA time series. In view
of solar cycling the model permits one to predict a time series of
soft X-ray solar flares, when the solar activity will be again 
near its maximum in 2010-2014 years. While we do not
claim that this model provides the only possible explanation, it
does provide a rigorous statistical picture of the expected observations of
X -ray solar flares.

\acknowledgements

A.A.S. is grateful to the Institute of Physics and the Hugo
Steinhaus Center for Stochastic Methods for pleasant hospitality
during his visit in Wroc{\l}aw University of Technology.
The GOES X-ray light curve was made available courtesy of the NOAA
Space Environment Center, Boulder, CO.

\clearpage

\clearpage

\clearpage

\clearpage

\clearpage

\clearpage

\clearpage

\clearpage

\begin{table}
\begin{center}
\caption{Values of the FIRT, VR and RS estimators for the original time series and the
shuffled (surrogate) solar flare data.}\label{tab1}
\begin{tabular}{cccc}
\\
\tableline\tableline Data set & $H_{FIRT}$ & $H_{VR}$ & $d_{RS}$\\ \tableline
\multicolumn{4}{c}{Original time series} \\ \tableline Solar flares & $1.1424$ &
$1.0665$ & $0.2408$\\ \tableline \multicolumn{4}{c}{Surrogate data}
\\ \tableline Solar flares & $0.8452$ & $0.7722$ & $0.0507$\\ \tableline
\end{tabular}
\end{center}
\end{table}

\end{document}